\documentstyle[amsfonts,epsfig,preprint,aps]{revtex}
\tightenlines
\begin{document}
\draft
\preprint{ }

\title{A self-consistent Ornstein-Zernike approximation for the Edwards-Anderson spin glass model\footnote{
This paper is dedicated to George Stell.}.}
\author{E. Kierlik, M. L. Rosinberg, and G.Tarjus}
\address{Laboratoire de Physique Th\'eorique des Liquides \cite{AAAuth},
Universit\'e Pierre et Marie Curie,\\ 
4 Place Jussieu, 75252 Paris Cedex 05, France}

\maketitle

\begin{abstract}

We propose a self-consistent Ornstein-Zernike approximation
for studying the Edwards-Anderson spin glass model. By 
performing two Legendre transforms in replica space, we introduce a Gibbs free energy depending
on both the magnetizations and the overlap order parameters. The correlation functions and the thermodynamics are then
obtained from the solution of a set of coupled partial differential equations. The approximation becomes
exact in the limit of infinite dimension and it provides a potential route for studying the stability of the high-temperature 
phase against replica-symmetry breaking fluctuations in finite dimensions. 
As a first step, we present the numerical predictions for the freezing temperature and the zero-field thermodynamic properties above freezing
as a function of dimensionality.
\end{abstract}

\pacs{{\bf Key words}: Disordered systems, spin glasses, Ornstein-Zernike equations.}

\newpage

\def\be{\begin{equation}}
\def\ee{\end{equation}}
\def\bea{\begin{eqnarray}}
\def\eea{\end{eqnarray}}
\def\bit{\begin{itemize}}
\def\eit{\end{itemize}}

\section{Introduction}

This paper is the third one in a series devoted to the application of the self-consistent Ornstein-Zernike approximation (SCOZA) to classical
spin systems with quenched disorder. The SCOZA has been formulated some time ago as a theory for simple fluid
and lattice-gas systems \cite{HS77}. It is based on an Ornstein-Zernike (OZ) approximation for the direct correlation function $c(r)$ that,
by construction, enforces consistency between the different routes that give the thermodynamic potentials in terms of the pair correlation
functions. In the language of magnetic systems, this means that  the same Gibbs free energy
${\cal G}(m,\beta)$ is obtained whether one integrates the susceptibility $\chi$ with respect to the magnetization $m$ or the internal energy $U$ with respect 
to the inverse temperature $\beta=1/k_BT$. In the case of the spin $1/2$ Ising model with ferromagnetic nearest-neighbor interactions, this 
self-consistency requirement is embodied in a diffusive-like partial differential equation with $m$ and $\beta$ as independent variables.
This equation could be solved only recently for the various three-dimensional
lattices \cite{DS1996}, showing that the SCOZA predicts  all thermodynamic and structural properties with great accuracy, even in the close vicinity of the critical point \cite{PSD1998}. As the spatial dimension $d$ goes to infinity, the SCOZA becomes exact and identifies with mean-field theory. This 
approach can thus be viewed as an approximate but non-perturbative way of taking into account thermal fluctuations in finite dimensions. 
Although this is an OZ theory which does not handle correctly long-range critical fluctuations, the asymptotic 
critical  exponents are non-classical. For the 3-d Ising model, they have
the spherical-model values in a very narrow region above the critical temperature, but are much accurate along 
the magnetization curve ($\beta=0.35$) \cite{PSD1998,HPS1999}.

In the preceding papers of this series \cite{KRT1997,KRT1999}, we used the replica method to generalize this approach to disordered spin systems.
This allowed us to obtain an accurate description of the dependence of the critical 
temperature on dilution in the 3-d site-diluted Ising model \cite{KRT1997} and to study the influence of the random-field
distribution on the phase diagram of the random-field Ising model (RFIM) for $d > 4$ \cite{KRT1999}. In both cases, possible replica-symmetry breaking (RSB) effects were ignored. In this work, we present a first application of the SCOZA to the Edwards-Anderson (EA) Ising spin glass
model \cite{EA1975} whose low-temperature properties remain a subject of controversies after nearly twenty-five years of intense activity.
Whereas Parisi's mean-field theory \cite{P1979} is generally accepted as the exact solution of
the infinite-ranged Sherrington-Kirkpatrick (SK) model \cite{SK1975,BY1986}, there is yet no 
consensus on whether or not the RSB scenario associated with the appearance of multiple equilibrium states at low temperature
survives in realistic, short-ranged, finite-dimensional models (for recent reviews of this problem,
see \cite{MPR1998,DKT1998}). 
A relevant open question is the existence of a phase transition in presence of an external magnetic field.
Whereas the scaling approach based on the droplet model \cite{McM1984,BM1985,FH1988} predicts that a non-zero magnetic field destroys the
spin glass phase, the mean-field picture suggests that there is still an Almeida-Thouless (AT) line \cite{AT1978} separating the 
spin glass phase from the paramagnetic phase.

Although the SCOZA is based on a simple OZ ansatz for the correlation functions (so that the critical exponent $\eta$ is zero, a rather crude
approximation for the 3-d EA model according to numerical simulations \cite{BPC1996}), it may be sufficiently accurate 
to give useful indications about the stability of the high temperature phase against RSB fluctuations. It is worth stressing that the SCOZA 
becomes exact
when $d \rightarrow \infty $, so that one recovers in this limit the behavior of the SK model.
Our purpose is thus to use this non-perturbative approach
to generalize the AT analysis to finite dimensions, i.e. to determine the locus of singularities of the spin glass
susceptibility $\chi_{SG}$ working from the replica-symmetric region. This amounts to study the eigenvalues of the wave-vector-dependent 
inverse susceptibility matrix at ${\bf k}={\bf 0}$. As the inverse susceptibility matrix is just the matrix of direct correlation functions
generated by a Legendre transformed Gibbs free energy, our program is the following: derive the appropriate OZ equations,
assume an OZ form for the direct correlation functions (which means, in the present case, truncate them at nearest-neighbor
separation), invert the OZ equations, and use all existing self-consistency requirements and exact relations (for instance, the values of the
pair distribution functions at zero separation) to derive a set of partial differential equations
 whose solution will provide the phase boundaries 
in zero field and the AT lines. (Note that the SCOZA can also be combined with the Parisi RSB scheme; the main obstacle is the tractability
of the calculation that requires inverting ultrametric matrices as done in Ref. \cite{DKT1998}.)
This program, however, is not fully completed in the present paper which is only devoted to the calculation of the freezing temperature
$T_f$ and of the thermodynamic properties at and above $T_f$ as a function of dimensionality.
The stability analysis, which requires a more difficult numerical computation, will
be presented later. The paper is organized as follows. In section 2, we introduce the model and we perform the double Legendre
transform that allows one to introduce the direct correlation functions. The OZ equations in replica space are then solved in the case of 
replica symmetry. In section 3, we define an OZ approximation for the direct correlation functions which, at the lowest level, 
coincides with standard mean field theory. We also derive the exact core conditions and self-consistency relations. 
In section 4, this formalism is used to derive the SCOZA equations in the case of zero field. 
Numerical results are presented in section 5.

\section{The replica-symmetric Ornstein-Zernike equations}

We consider an EA spin-glass Hamiltonian for $N$ spins $\sigma_i=\pm 1$ on a d-dimensional hypercubic lattice 
\be
{\cal H} = - \sum_{<ij>}J_{ij} \sigma_{i}\sigma_{j} -H \sum_i \sigma_i 
\ee
where the first sum runs over all nearest neighbor pairs and the couplings $J_{ij}$  are 
independent Gaussian variables with mean $J_0$ and variance $J^2$; $H$ is an external magnetic field.

Introducing as usual $n$ replicas $\sigma_i^a$ ($a=1,...,n$) of the original spin variables and performing the average over
disorder, one finds that the quenched free energy is given by 

\be
{\cal F}=- \frac{1}{\beta}\lim_{n \rightarrow 0} \frac {1}{n}[\exp(-\beta {\cal F}_{n})-1] 
=- \frac{1}{\beta} \lim_{n \rightarrow 0} \frac {1}{n}[Tr \exp(-\beta {\cal H}_{n})-1]  
\ee
where ${\cal H}_{n}$ is a temperature-dependent effective Hamiltonian

\be
{\cal H}_{n}= -\sum_{<ij>} [J_0\sum_{a}\sigma_{i}^a\sigma_{j}^a 
+\frac{\beta J^2}{2} \sum_{a,b}\sigma_{i}^a\sigma_{i}^b\sigma_{j}^a\sigma_{j}^b]-H \sum_{i}\sum_{a}\sigma_i ^a \  ,
\ee
and the trace in Eq. (2) is taken over the $\sigma_i^a $'s. 
It is also useful to introduce a fictitious field $\Delta$ which couples to the quantity  
$-1/2 \sum_i\sum_{a,b}\sigma_i^a\sigma_i^b$. When $\Delta >0$, $k_BT\Delta$ can be interpreted as the variance of a 
Gaussian random field but, in the present work, we are mainly interested in the limit $\Delta =0$ (see Ref. \cite{SNA1994} for
a study of the SK model in the presence of a Gaussian random field)
We treat $H$ and $\Delta$ as sources that we extend to site- and replica-dependent values $H_i^a$ and $\Delta_i^{ab}$ 
in order to generate the correlation functions. We thus consider the more general Hamiltonian

\be
{\cal H}_{n}= -\sum_{<ij>}\ [ J_0 \sum_{a}\sigma_{i}^a\sigma_{j}^a 
+\beta J^2\sum_{a<b}\sigma_{i}^a\sigma_{i}^b\sigma_{j}^a\sigma_{j}^b]-\sum_{i}\ [\sum_a  H_i^a\sigma_i ^a
+\sum_{a<b} \Delta_i^{ab}\sigma_{i}^a\sigma_{i}^b]
\ee
where we have restricted $\Delta_i^{ab}$ to $a<b$ with the condition $\Delta_i^{ab}=\Delta_i^{ba}$  and omitted the 
constant contribution $-n N [c\beta J^2/4 +\Delta/2]$ of
the diagonal terms ($c$ is the coordination number of the lattice). 
The averaged magnetization $m_i ^a$  and overlap parameters $q_i^{ab}$ are given  by
 
\be
m_i ^a=-\frac{\partial { {\cal F}_n}}{\partial  H_i^a}=<\sigma_i^a>_n
\ee
and
\be
q_i^{ab}=-\frac{\partial {{\cal F}_n}}{\partial  \Delta_i^{ab}}=<\sigma_i^a\sigma_i^b>_n \ 
\ee
respectively, where $<...>_{n}$ denotes the replica thermal average.
At the end, we shall take the limit $H_i^{a}\rightarrow H$ and $\Delta_i^{ab}\rightarrow \Delta$.
Then, in the limit $n \rightarrow 0$, one has $m_i^a \rightarrow m$, and,
if replica-symmetry holds, $q_i^{ab}\rightarrow q$ (which is the also the usual Edwards-Anderson order parameter $q_{EA}$ 
when $\Delta=0$). Accordingly, one has 
$\partial ({ {\cal F}}/N)/\partial  H=-m$ and $\partial ({{\cal F}}/N)/\partial  \Delta=\frac{1}{2}(q-1)$ (the $-1/2$ comes from the constant 
contribution omitted in Eq. (4)).

The second partial derivatives with respect to $H_i^a$ and $\Delta_i^{ab}$ define the matrix of (connected) 
correlation functions in replica space

\be
G_{ij}^{ab}=- \frac{\partial ^2 {\tilde {\cal F}}_n}{\partial \tilde H_i^a \partial \tilde H_j^b} 
=<\sigma_i^a\sigma_j^b>_n-<\sigma_i^a>_n<\sigma_j^b>_n
\ee
\be
G_{ij}^{ab,cd}= -\frac{\partial ^2 {\tilde{\cal F}}_n}{\partial \tilde \Delta_i^{ab} \partial \tilde \Delta_j^{cd}}
= <\sigma_i^a\sigma_i^b\sigma_j^c\sigma_j^d>_n-<\sigma_i^a\sigma_i^b>_n<\sigma_j^c\sigma_j^d>_n \ \ (a<b, c<d) 
\ee
\be
G_{ij}^{a,bc}=-\frac{\partial ^2 {\tilde{\cal F}}_n}{\partial \tilde H_i^a \partial \tilde \Delta_j^{bc}}
= <\sigma_i^a\sigma_j^b\sigma_j^c>_n-<\sigma_i^a>_n<\sigma_j^b\sigma_j^c>_n \ \ (b<c) 
\ee
and similarly for $G_{ij}^{ab,c}$ (${ \tilde{\cal F}}_n= \beta {\cal F}_n$, $\tilde H= \beta H$, and $\tilde \Delta= \beta \Delta$).

We now perform a double Legendre transform that takes the fields $H_i^a$ and $\Delta_{ij}^{ab}$ into $m_i^a$ and $q_i^{ab}$, respectively. 
This defines the Gibbs free energy

\be
 {\cal G}_{n}={\cal F}_{n}+\sum_i\sum_aH_i^a m_i^a+\sum_i\sum_{a<b}\Delta_i^{ab} q_i^{ab}
\ee
which satisfies $ H_i^a=\partial {{\cal G}}_n/\partial m_i^a$ and $ \Delta_i^{ab}=\partial {{\cal G}}_n/\partial q_i^{ab}$.
 Accordingly, in the limit $n \rightarrow 0$, $m_i^a \rightarrow m$, and $q_i^{ab}\rightarrow q$, 
the Gibbs free energy ${  {\cal G}}(m,q,\beta)$  satisfies
\be
\frac{\partial ({  {\cal G}}/N)}{\partial m}= H
\ee
and 
\be
\frac{\partial ({  {\cal G}}/N)}{\partial q}=-\frac{1}{2} \Delta .
\ee
${\cal G}_{n}$ is the generating functional of the direct correlation functions (or proper vertices in field-theoretical language)
\be
C_{ij}^{ab}= \frac{\partial ^2 {\tilde{\cal G}}_n}{\partial m_i^a \partial m_j^b}
\ee
\be
C_{ij}^{ab,cd}= \frac{\partial ^2 {\tilde{\cal G}}_n}{\partial q_i^{ab} \partial q_j^{cd}}\ \ (a<b, c<d) 
\ee
\be
C_{ij}^{a,bc}=\frac{ \partial ^2 {\tilde{\cal G}}_n}{\partial m_i^a \partial q_j^{bc}}\ \ (b<c) 
\ee
and similarly for $C_{ij}^{ab,c}$ (${\tilde {\cal G}}_{n}=\beta {\cal G}_n$).

The matrix ${\underline{\underline {\bf C}}}$ is the inverse of the matrix ${\underline{\underline {\bf G}}}$ and this
defines a set of ``Ornstein-Zernike equations''. 
In the limit of uniform fields, one has in Fourier space

\be
\sum_c \hat{C}^{ac}({\bf k})\hat{G}^{cb}({\bf k})+\sum_{c<d} \hat{C}^{a,cd}({\bf k})\hat{G}^{cd,b}({\bf k})=\delta_{a,b} 
\ee
and
\be
\sum_e \hat{C}^{ab,e}({\bf k})\hat{G}^{e,cd}({\bf k})+\sum_{e<f} \hat{C}^{ab,ef}({\bf k})\hat{G}^{ef,cd}({\bf k})=\delta_{a,c} \ \delta_{b,d} \ .
\ee

These equations are the starting point of our study and the first task is to solve them,
 i.e. to express the propagators ${\hat G}$'s in terms of the direct correlations functions ${\hat C}$'s (or, conversely, the 
${\hat C}$'s in terms of the ${\hat G}$'s) for general values of $n$ and then take the limit $n \rightarrow 0$.
This is easy when replica-symmetry holds since $\hat{{\underline{\underline {\bf G}}}}$ and $\hat{{\underline{\underline {\bf C}}}}$ 
have the same structure as the Hessian or stability matrix of the SK model which has been analyzed by de Almeida and Thouless \cite{AT1978}
(on the other hand, the problem becomes highly non-trivial when continuous RSB {\it \`a la} Parisi occurs, since one has to cope with the 
inversion of ultrametric matrices , see. e.g. \cite{DKT1998}).
There are seven different types of matrix elements, $\hat{G}^{aa}$, $\hat{G}^{ab}$, 
$\hat{G}^{ab,ab}$, $\hat{G}^{ab,ac}$, $\hat{G}^{ab,cd}$, $\hat{G}^{a,ab}=\hat{G}^{ab,a}$, $\hat{G}^{c,ab}=\hat{G}^{ab,c}$ (resp. $\hat{C}^{aa}$, $\hat{C}^{ab}$,...etc) and the two matrices of order $n(n+1)/2$  can be easily block-diagonalized.
There are three subspaces of dimensions $2$, $2(n-1)$ and $n(n-3)/2$ which, following the standard terminology 
\cite{DKT1998,BM1979}, we call respectively L (for longitudinal), A (for anomalous) and R (for replicon). 
The block-diagonalized matrices then contain a single $2 \times 2$ L-block, $n-1$ identical $2 \times 2$ A-blocks, 
and $n(n-3)/2$ identical R-eigenvalues.
In the new representation, the Ornstein-Zernike equations, hereafter called the 
replica-symmetric Ornstein-Zernike (RSOZ) equations, break up into three sets of equations. The first set reads
\be
 \hat{{\underline{\underline  {\bf G}}}}^L({\bf k})= \hat{{\underline{\underline  {\bf C}}}}^L({\bf k})^{-1}
\ee
with 

\be
\hat{{\underline{\underline {\bf C}}}}^L= \left(\begin{array}{ccr}
\hat{C}^{aa}+(n-1)\hat{C}^{ab}\ \ &(n-1)\hat{C}^{a,ab}+\frac{(n-1)(n-2)}{2}\hat{C}^{a,bc}\\
2\hat{C}^{a,ab}+(n-2)\hat{C}^{a,bc}\ \ &\hat{C}^{ab,ab}+2(n-2)\hat{C}^{ab,ac}+\frac{(n-2)(n-3)}{2}\hat{C}^{ab,cd}
\end{array} \right) 
\ee
Similar expressions hold for the elements of ${\underline{\underline {\bf \hat{G}}}}^L$. The second set reads
\be
\hat{{\underline{\underline  {\bf G}}}}^A({\bf k})= \hat{{\underline{\underline  {\bf C}}}}^A({\bf k})^{-1}
\ee
with
\be
\hat{{\underline{\underline {\bf C}}}}^A= \left(\begin{array}{ccr}
\hat{C}^{aa}-\hat{C}^{ab}\ \ &(n-1)(\hat{C}^{a,ab}-\hat{C}^{a,bc})\\
\frac{n-2}{n-1}(\hat{C}^{a,ab}-\hat{C}^{a,bc})\ \ &\hat{C}^{ab,ab}+(n-4)\hat{C}^{ab,ac}-(n-3)\hat{C}^{ab,cd}
\end{array} \right) 
\ee
and similar expressions for ${\underline{\underline {\bf \hat{G}}}}^A$. The last equation corresponds to the replicon sector,

\be
\hat{G}^R({\bf k})= \hat{C}^R({\bf k})^{-1}
\ee
with 
\be
\hat{C}^R=\hat{C}^{ab,ab}-2\hat{C}^{ab,ac}+\hat{C}^{ab,cd}
\ee
and a similar expression for $\hat{G}^R$. 

In the limit $n \rightarrow 0$, Eq (19) readily yields

\begin{mathletters}
\bea
\hat{C}^L_{11}&=&\hat{C}^{aa}-\hat{C}^{ab}\\
\hat{C}^L_{12}&=&\hat{C}^{a,bc}-\hat{C}^{a,ab}\\
\hat{C}^L_{21}&=&2(\hat{C}^{a,ab}-\hat{C}^{a,bc})=-2\hat{C}^L_{12}\\
\hat{C}^L_{22}&=&\hat{C}^{ab,ab}-4\hat{C}^{ab,ac}+3\hat{C}^{ab,cd}\ ,
\eea
\end{mathletters}
and similarly for the four elements of the matrix $\hat{{\underline{\underline {\bf G }}}}^L$. They can be deduced from the elements of
$\hat{{\underline{\underline {\bf C}}}}^L$ by using Eq. (18) (for simplicity, in these expressions
and in the following, we keep the same notations for $\hat{G}^L$, $\hat{C}^L$, $\hat{G}^{aa}$, etc... although the 
limit  $n \rightarrow 0$ has been taken).

When $n  \rightarrow 0$, we see from Eqs. (19) and (21) that
$ \hat{{\underline{\underline {\bf C}}}}^A  \rightarrow \hat{{\underline{\underline {\bf C}}}}^L$ (resp. 
$ \hat{{\underline{\underline {\bf G}}}}^A  \rightarrow \hat{{\underline{\underline {\bf G}}}}^L$). But the difference
$ \hat{{\underline{\underline {\bf C}}}}^A - \hat{{\underline{\underline {\bf C}}}}^L$ is of order $n$, which yields 
\bea
\lim_{n \rightarrow 0}\frac{1}{n}( \hat{{\underline{\underline {\bf C}}}}^L-\hat{{\underline{\underline {\bf C}}}}^A)&=&
 \left(\begin{array}{ccr}
\hat{C}^{ab}\ \ &-\frac{1}{2}\hat{C}^{a,bc}\\
2\hat{C}^{a,bc}-\hat{C}^{a,ab}\ \ &\hat{C}^{ab,ac}-\frac{3}{2}\hat{C}^{ab,cd}
\end{array} \right) \\
&=& (\hat{{\underline{\underline {\bf G}}}}^L)^{-1}. \left(\begin{array}{ccr}
-\hat{G}^{ab}\ \ &\frac{1}{2}\hat{G}^{a,bc}\nonumber\\
-2\hat{G}^{a,bc}+\hat{G}^{a,ab}\ \ &-\hat{G}^{ab,ac}+\frac{3}{2}\hat{G}^{ab,cd}
\end{array} \right) . \  (\hat{{\underline{\underline {\bf G}}}}^L)^{-1}
\eea

This set of equations allows to express all the correlation functions $\hat{C}$'s in terms of the $\hat{G}$'s (and conversely) 
in the limit $n \rightarrow 0$.
The values of the direct correlation functions at ${\bf k}= {\bf 0}$ in the longitudinal sector 
are directly related to the various susceptibilities. One  has

\begin{mathletters}
\bea
\frac{\partial ^2 ({\tilde  {\cal G}}/N)}{\partial m^2}=\frac{\partial \tilde H}{\partial m}=\hat{C}^L_{11}({\bf k}={\bf 0})
\eea
\bea
\frac{\partial ^2 ({\tilde  {\cal G}}/N)}{\partial m \partial q}=
\frac{\partial \tilde H}{\partial q}=-\frac{1}{2} \frac{\partial \tilde \Delta}{\partial m}=\hat{C}^L_{12}({\bf k}={\bf 0}) 
\eea
\bea
\frac{\partial ^2 ({\tilde  {\cal G}}/N)}{\partial q^2}
=-\frac{1}{2}\frac{\partial \tilde \Delta}{\partial q}=-\frac{1}{2}\hat{C}^L_{22}({\bf k}={\bf 0}) \ . 
\eea
\end{mathletters}
In particular, when $H=\Delta=0$, $\hat{C}^L_{11}({\bf k}={\bf 0})=\beta \chi^{-1}$, the inverse of the magnetic susceptibility, whose divergence 
signals the occurence of the paramagnetic to ferromagnetic transition. More generally, 
in the approximate theory that is discussed below, the vanishing of the determinant of ${\underline{\underline {\bf C}}}^L$
defines spinodal lines.
On the other hand, the stability limit of the replica-symmetric solution (i.e. the AT lines) is signaled by the vanishing of 
$\hat{C}^R({\bf k}={\bf 0})$. In zero fields, one has 
\be
\hat{C}^R({\bf k}={\bf 0})=\frac{1}{\hat{G}^R({\bf k}={\bf 0})}=\chi_{SG}^{-1}
\ee
where $\chi_{SG}=1/N \sum_{i,j}[(<\sigma_i \sigma_j>_T-<\sigma_i>_T<\sigma_j>_T)^2]_{av}$,
is the spin-glass susceptibility \cite{BY1986}. 

It is also useful to note the consequences of the symmetry properties of the Hamiltonian ${\cal H}_n$ described by Eq. (4) for 
the correlation functions. 
${\cal H}_n$ is invariant under the global transformation $\{\sigma_i^a\},\{H_i^a\} \rightarrow \{-\sigma_i^a\},\{-H_i^a\}$ for all spins in all replicas
while leaving $\{\Delta_i^{ab}\}$ unchanged. In consequence, averaged quantities involving an odd number of spins are zero 
when $H=m=0$ (even when $\Delta \neq 0$). 
In particular, $G^{a,ab}({\bf r})=G^{a,bc}({\bf r})=C^{a,ab}({\bf r})=C^{a,bc}({\bf r})=0$. This implies that $G^L_{12}({\bf r})=G^L_{21}({\bf r})
=C^L_{12}({\bf r})=C^L_{21}({\bf r})=0$ and the RSOZ equations involving the 2-replicas and 4-replicas correlation functions, Eqs. (16) and 
(17), decouple. 
${\cal H}_n$ is also invariant in the global transformation
$\{\sigma_i^a\},\{H_i^a\},\{\Delta_i^{ab}\} \rightarrow \{-\sigma_i^a\},\{-H_i^a\},\{-\Delta_i^{ab}\}$ for all spins in a {\it single} replica $a$. 
This implies that
$G^{ab}({\bf r})=G^{ab,ac}({\bf r})=G^{ab,cd}({\bf r})=C^{ab}({\bf r})=C^{ab,ac}({\bf r})=C^{ab,cd}({\bf r})=0$ in the paramagnetic phase
when $H=\Delta=0$. Finally, when $J_0=0$, ${\cal H}_n$ is also invariant in the local transformation
$\{\sigma_i^a\},\{H_i^a\} \rightarrow \{-\sigma_i^a\},\{-H_i^a\}$ for a  {\it single} spin $i$ in all replicas
while leaving $\{\Delta_i^{ab}\}$ unchanged. This implies that  $m=0$ and that $G^ {aa}({\bf r})$ and $G^ {ab}({\bf r})$ are local when
$H=0$ (even when $\Delta \neq 0$). More precisely, one has $G^ {aa}({\bf r})=\delta_{{\bf r},{\bf 0}}$ and 
$G^ {ab}({\bf r})=q\delta_{{\bf r},{\bf 0}}$. This implies that $\chi=\beta \hat{G}^L_{11}({\bf k}={\bf 0})=\beta (1-q)$ in this case, 
as is well known \cite{BY1986} (the first equality follows from the fact that $C^L_{12}({\bf r})=C^L_{21}({\bf r})=0$).

\section{Ornstein-Zernike approximation and exact relations}

So far, all equations are exact provided that replica symmetry is unbroken. We now introduce an OZ approximation for
the direct correlation functions. Since the interactions in the effective Hamiltonian are restricted to nearest-neighbors, 
we assume that the seven distinct C's are non zero only at ${\bf r}=0$ and ${\bf r}={\bf e}$, where ${\bf e}$ is a
vector from the origin to one of its nearest-neighbors. On the other hand, their dependence on $m$, $q$ and $\beta$ is not
given {\it a priori}. We thus write that

\bea
C^{x}({\bf r})=c_0^{x}(m,q,\beta) \delta_{{\bf r},{\bf 0}}+c_1^{x}(m,q,\beta) \delta_{{\bf r},{\bf e}}
\eea
where $x=aa,ab, aab$, etc...The whole problem lies in the determination of the $c_0$'s and the $c_1$'s.

When $J_0=J=0$ (or at infinite temperature with $\tilde H$ and $\tilde \Delta$ fixed), the correlation functions are local (i.e. the $c_1$'s are zero) and can be calculated by 
averaging directly over the disorder distribution. Hereafter, we call this system the reference system. In terms of $m$ and $q$,
one finds that 
$G^{aa}_{ref}({\bf r})=(1-m^2)\delta_{{\bf r},{\bf 0}}$, 
$G^{ab}_{ref}({\bf r})=(q-m^2)\delta_{{\bf r},{\bf 0}}$,
$G^{a,ab}_{ref}({\bf r})=m(1-q)\delta_{{\bf r},{\bf 0}}$, 
$G^{a,bc}_{ref}({\bf r})=(t-mq)\delta_{{\bf r},{\bf 0}}$, $G^{ab,ab}_{ref}({\bf r})=(1-q^2)\delta_{{\bf r},{\bf 0}}$,
$G^{ab,ac}_{ref}({\bf r})=q(1-q)\delta_{{\bf r},{\bf 0}}$, $G^{ab,cd}_{ref}({\bf r})=(r -q^2)\delta_{{\bf r},{\bf 0}}$,
with
\begin{mathletters}
\bea
m=\int_{-\infty}^{+\infty}{\cal D}x \tanh(\tilde H_{ref}+x\tilde \Delta_{ref}^{1/2})
\eea
\bea
q=\int_{-\infty}^{+\infty} {\cal D}x \tanh^2(\tilde H_{ref}+x\tilde \Delta_{ref}^{1/2})
\eea
\bea
t=\int_{-\infty}^{+\infty}{\cal D}x  \tanh^3(\tilde H_{ref}+x\tilde \Delta_{ref}^{1/2})
\eea
\bea
r=\int_{-\infty}^{+\infty}{\cal D}x \tanh^4(\tilde H_{ref}+x\tilde \Delta_{ref}^{1/2}) \ .
\eea
\end{mathletters}
where ${\cal D}x=1/\sqrt(2 \pi) dx \exp(-x^2/2) $. Here, $H_{ref}$ and $\Delta_{ref}$ must be considered as auxiliary field variables that can 
be eliminated from Eqs. (29a) and (29b) to express $t$ and $r$ as functions of $m$ and $q$, for instance as infinite double series,
\begin{mathletters}
\bea
t(m,q)=3mq-2m(m^2+3q^2)+6mq(2m^2+3q^2)-6m(m^4+10m^2q^2+13q^4)+...
\eea
\bea
r(m,q)=3q^2-8q^3-2(m^4-16q^4)+24q(m^4-7q^4)+...
\eea
\end{mathletters}
The corresponding direct correlation functions are obtained from the RSOZ equations, Eqs. (20-27) (inverting the r\^{o}le of 
${\underline{\underline {\bf G}}}$ and ${\underline{\underline {\bf C}}}$). Their expressions are given in the Appendix.

Now, the simplest approximation for the direct correlation functions at finite temperature is the Random Phase Approximation (RPA). 
It consists in setting the $c_1$'s equal to $-\beta$ times the corresponding interactions in the Hamiltonian. Here, this gives

\be
\hat{C}^{aa}_{RPA}({\bf k})=c^{aa}_{0,ref}(m,q,\beta)- \tilde J_0 \hat{\lambda}({\bf k})
\ee
\be
\hat{C}^{ab,ab}_{RPA}({\bf k})=c^{ab,ab}_{0,ref}(m,q,\beta)- \tilde J^2\hat{\lambda}({\bf k})
\ee
where $\tilde J_0=c\beta J_0$, $\tilde J=c^{1/2}\beta J$, and $\hat{\lambda}({\bf k})=1/c \sum_{{\bf e}}\exp(i {\bf k}.{\bf e})$ is
the characteristic function of the lattice. All other $C$'s remain equal to the corresponding $C$'s of the reference system, 
$\hat{C}^{ab}_{RPA}({\bf k})=c_{0,ref}^{ab}(m,q,\beta)$, $\hat{C}^{a,ab}_{RPA}({\bf k})=c_{0,ref}^{a,ab}(m,q,\beta)$, etc...

Using Eqs. (A1) and (A2), one then finds

\begin{mathletters}
\be
\hat {C}^L_{11,RPA}({\bf k})=(1-4q+3r)D_{ref}- \tilde J_0 \hat{\lambda}({\bf k})
\ee
\be
\hat {C}^L_{22,RPA}({\bf k})=(1-q)D_{ref}- \tilde J^2\hat{\lambda}({\bf k})
\ee
\be
\hat {C}^L_{12,RPA}({\bf k})=-\frac{1}{2}\hat {C}^L_{21,RPA}({\bf k})=(m-t)D_{ref}
\ee
\end{mathletters}
so 
\begin{mathletters}
\be
\frac{\partial \tilde H}{\partial m}=(1-4q+3r)D_{ref}- \tilde J_0 
\ee
\be
\frac{\partial \tilde \Delta}{\partial q}=(1-q)D_{ref}- \tilde J^2
\ee
\be
\frac{\partial \tilde H}{\partial q}=-\frac{1}{2} \frac{\partial \tilde \Delta}{\partial m}=(m-t)D_{ref} \ .
\ee
\end{mathletters}
It is straightforward  to check that these results are also obtained by differentiating the following expression of the
free energy

\bea
{\tilde{\cal F}}/N =-\frac{\tilde J^2}{4}(1-q)^2+\frac{\tilde J_0}{2}m^2-\int_{-\infty}^{+\infty}{\cal D}x \ \ln [2\cosh (\tilde H+\tilde J_0m+x(\tilde \Delta+\tilde J^2q)^{1/2}]
\eea
where $m$ and $q$ are given by Eqs. (29a) and (29b)   
with $\tilde H_{ref}$ replaced by $\tilde H +\tilde J_0m$ and $\tilde \Delta_{ref}$ by $\tilde \Delta+\tilde J^2q$ 
(the same modification must be performed, of course, in the definitions of $t$ and $r$). 
Eq. (35) is just the replica-symmetric solution of the SK model  in the presence of a Gaussian random field \cite{SNA1994}. 
Therefore, the RPA is equivalent to mean-field 
theory when the Gibbs free energy is obtained by integration of 
the $H$- or $\Delta$-susceptibility. (This statement is actually valid irrespective of the assumption of replica symmetry.) 
In particular, the instability with respect to RSB is given by the vanishing of

\bea
\hat {C}^R_{RPA}({\bf k}={\bf 0})=\hat {C}^R_{ref}[1-z^{RPA}_R]
\eea
where
\be
z^{RPA}_R=(1-2q+r) \tilde J^2 \ .
\ee
When $\Delta=0$, the condition $z^{RPA}_R=1$ yields the usual AT lines  \cite{AT1978}.

Mean field theory becomes exact when $d \rightarrow \infty$. On the other hand, as is well-known in another context (see e.g., \cite{HMcD1976}), 
the RPA in finite dimensions predicts pair distribution functions that do not satisfy the proper sum rules in the ``core'', i.e. when 
${\bf r}={\bf 0}$. Here, the exact core conditions are obtained by using the hard spin condition $\sigma_i=\pm 1$ and the definition of
$q$. One finds in the replica-symmetric case

\begin{mathletters}
\be
G^{aa}({\bf r}={\bf 0})=1-m^2
\ee
\be
G^{ab}({\bf r}={\bf 0})=q-m^2
\ee
\be
G^{a,ab}({\bf r}={\bf 0})=m(1-q)
\ee
\be
G^{ab,ab}({\bf r}={\bf 0})=1-q^2
\ee
\be
G^{ab,ac}({\bf r}={\bf 0})=q(1-q) \ .
\ee
\end{mathletters}
It can be checked that  these relations are not satisfied by the RPA. This disease may be cured by adjusting the values of the 
corresponding direct correlation functions at ${\bf r}={\bf 0}$, i.e. the values of  $c_0^{aa},c_0^{ab}, c_0^{a,ab},c_0^{ab,ab}$ 
and $c_0^{ab,ac}$. In liquid-state theory, such an
approximation is called the Optimized Random Phase Approximation (ORPA) \cite{AC1972} and is closely related to
the mean-spherical approximation. We have seen in Ref. \cite{KRT1999} in the case of the RFIM that going from the RPA to the ORPA
improves the predictions for non-universal quantities such as the critical temperature. It 
also modifies the critical behavior because of a subtle interplay between small-$r$ and large-$r$ correlations \cite{S1969}. 
On the other hand, the ORPA (like the RPA) is not thermodynamically self-consistent as different Gibbs free energies
are obtained depending on the route that is chosen for calculating ${\cal G}$ from the pair correlation functions.
 In order to get a self-consistent theory we have also to adjust the values of the direct correlation 
functions at ${\bf r}={\bf e}$, i.e. the values of the $c_1$'s.

In the EA model, self-consistency is embodied in three Maxwell relations that can be obtained by considering the 
variation of ${\cal G}$ as one varies  the control variables $J, m$ and $q$ independently while keeping the ratio $J_0/J$ fixed. 
From Eqs. (4) and (10), one finds in the replica-symmetric $n \rightarrow 0$ limit

\be
\frac{\partial ({\tilde{\cal G}}/N)}{\partial {\tilde J}^2}=-\frac{1}{4}[ \frac{{\tilde J}_0}{{\tilde J}^2}(G^{aa}({\bf r}={\bf e})+m^2) 
-(G^{ab,ab}({\bf r}={\bf e})+q^2-1)] \ .
\ee

Then, using Eqs. (26) for the second partial derivatives of ${\tilde{\cal G}}$ with respect to $m$ and $q$, one finds that the 
cross-derivatives satisfy 

\begin{mathletters}
\bea
\frac{\partial \hat C_{11}^L ({\bf k}={\bf 0})}{\partial {\tilde J}^2}=-\frac{{\tilde J}_0}{2 {\tilde J}^2} -\frac{1}{4}\frac{\partial ^2}
{\partial m^2}[\frac{{\tilde J}_0}{{\tilde J}^2} G^{aa}({\bf r}={\bf e}) - G^{ab,ab}({\bf r}={\bf e})] \ 
\eea
\bea
\frac{\partial \hat C_{12}^L ({\bf k}={\bf 0})}{\partial {\tilde J}^2}=-\frac{1}{4}\frac{\partial ^2}{\partial m \partial q} 
[\frac{{\tilde J}_0}{{\tilde J}^2} G^{aa}({\bf r}={\bf e}) - G^{ab,ab}({\bf r}={\bf e})]\ 
\eea
\bea
\frac{\partial \hat C_{22}^L ({\bf k}={\bf 0})}{\partial {\tilde J}^2}=-1 +\frac{1}{2}\frac{\partial ^2}{\partial q^2}
[\frac{{\tilde J}_0}{{\tilde J}^2} G^{aa}({\bf r}={\bf e}) - G^{ab,ab}({\bf r}={\bf e})] \  .
\eea
\end{mathletters}

Therefore, one has only five core conditions and three self-consistency relations available, whereas, from the OZ approximation, Eq. (28),
 there are forteen unknown functions $c_0$'s and $c_1$'s to determine. In order to solve the problem completely, one
 must thus introduce additional approximations.

\section{The case $H=J_0=0$}

As a first requirement, a sensible theory for the spin glass transition should yield reasonable predictions for the freezing temperature $T_f$ in zero field. 
We thus take $H=0$ in the following, and, to keep things simple, we only consider the case $J_0=0$. Then $m=0$ 
and the RSOZ equations, Eqs. (16) and (17), decouple, as noted earlier. Moreover, only the correlation functions depending on four
replica indices are relevant, so that only  two core conditions, Eqs. (38d) and (38e), and one self-consistency relation, Eq. (40c), come
into play. In order to have only three unknown state-dependent functions to determine, we shall assume 
that $c_{1}^{ab,ac}(q, \beta)=c_{1}^{ab,cd}(q, \beta)=0$ and $c_{0}^{ab,cd}(q, \beta)=c_{0,ref}^{ab,cd}(q)$. 
According to the ORPA philosophy, these two additional approximations seem quite natural since only two distinct replicas interact
in the Hamiltonian ${\cal H}_n$ and there is no core condition associated to $G^{ab,cd}$.
Our simplified OZ approximation for the direct correlation functions in Fourier space thus reads
\bea
\hat C^{ab,ab}({\bf k})&=&c_{0}^{ab,ab}(q,\beta)+c_1^{ab,ab}(q,\beta) \lambda({\bf k})\nonumber\\
\hat C^{ab,ac}({\bf k})&=&c_{0}^{ab,ac}(q,\beta) \nonumber\\
\hat C^{ab,cd}({\bf k})&=&c_{0,ref}^{ab,cd}(q)
\eea
Since the direct correlation functions have the same spatial structure as in the RPA, this theory
will reduce to mean field theory and become exact when $d \rightarrow \infty$.

It is now easy to calculate $G^{ab,ab}({\bf r})$ and $G^{ab,ac}({\bf r})$ so to use the two core conditions and
derive the SCOZA  partial differential equation (PDE). 
Introducing the auxiliary function 
$G^{D}({\bf r})= G^{ab,ac}({\bf r})-3/2 \ G^{ab,cd}({\bf r})$, one first notes that

\begin{mathletters}
\bea
G^{ab,ab}({\bf r})=-2[G^L_{22}({\bf r})-\frac{3}{2}G^R({\bf r})+G^{D}({\bf r})]
\eea
\bea
G^{ab,ac}({\bf r})=-\frac{3}{2}[G^L_{22}({\bf r})-G^R({\bf r})+\frac{4}{3}G^{D}({\bf r})]
\eea
\end{mathletters}
From Eqs. (41), one has

\be
\hat C^L_{22}({\bf k})=C^L_{0,22}[1-z \lambda({\bf k})]
\ee
where $C^L_{0,22}=c_{0}^{ab,ab}-4c_{0}^{ab,ac}+3c_{0,ref}^{ab,cd}$ and  $z=-c_1^{ab,ab}/C^L_{0,22}$.
Since the matrices ${\underline{\underline {\bf \hat{G}}}}^L$ and 
${\underline{\underline {\bf \hat{C}}}}^L$ are diagonal when $m=0$, one gets immediately

\be
G^L_{22}({\bf r})=G^L_{0,22}\ P({\bf r},z)
\ee
where $G^L_{0,22}=1/C^L_{0,22}$ and $P({\bf r},z)$ is the lattice Green's function defined by \cite{J1972}
\be
P({\bf r},z)=\frac{1}{(2\pi)^d}\int_{-\pi}^{\pi}d^d{\bf k}\frac{e^{i{\bf k}.{\bf r}}}{1-z\hat{\lambda}({\bf k})} \ .
\ee
Similarly, one has
\be
\hat C^R({\bf k})=C^R_{0}[1-z_R \lambda({\bf k})]
\ee
with $C^R_0=c_{0}^{ab,ab}-2c_{0}^{ab,ac}+c_{0,ref}^{ab,cd}$ and $z_R=-c_1^{ab,ab}/C^R_0$. Hence, from Eq. (22), 
\be
G^R({\bf r})=G^R_0\ P({\bf r},z_R)
\ee
with $G^R_0=1/C^R_0=(z_R/z) G^L_{0,22}$. Finally, from Eq. (25), one obtains

\be
G^{D}({\bf k})=\frac{G^{D}_0}{[1-z \lambda({\bf k})]^2}
\ee
with  $G^{D}_0=[3\ c_{0,ref}^{ab,cd}/2-c_{0}^{ab,ac}](G^L_{0,22})^2
=[c_{0,ref}^{ab,cd}+(1-z/z_R)/G^L_{0,22}](G^L_{0,22})^2/2$.
Hence,
\be
G^{D}({\bf r})=G^{D}_0\ \frac{\partial}{\partial z}[zP({\bf r},z)]
\ee
The core conditions, Eqs. (38d) and (38e), can then be used to express $G^L_{0,22}$ and $G^{D}_0$ in terms of 
$z$ and $z_R$. One finds
\be
G^L_{0,22}=\frac{2z(1-q)}{3z_R P(z_R)-zP(z)}
\ee
and 
\be
G^{D}_0=\frac{1-q}{zP'(z)+P(z)}[1-q-2z\frac{P(z)}{3z_R P(z_R)-zP(z)}]
\ee
where $P(z)\equiv P({\bf r}={\bf 0},z)$. Moreover, $z$ and $z_R$ are related via

\bea
(1-q)[3z_R P(z_R)-zP(z)]^2=&2&z[3z_R P(z_R)-zP(z)][P(z)+(1-\frac{z}{z_R})\frac{d}{dz}(zP(z))]\nonumber\\
&+& 4z^2(1-q) \frac{d}{dz}(zP(z))c_{0,ref}^{ab,cd}(q)
\eea
The self-consistency relation, Eq. (40c), can now be expressed as a PDE in the unknown function
$z(q,\lambda)$, where $\lambda \equiv \tilde J^2=c \beta^2 J^2$. One obtains

\be
\frac{\partial \hat C_{22}^L ({\bf k}={\bf 0})}{\partial \lambda}=-1 -\frac{1}{2}\frac{\partial ^2}{\partial q^2}G^{ab,ab}({\bf r}={\bf e})
\ee
with
\be
\hat C_{22}^L ({\bf k}={\bf 0})=\frac{3z_R P(z_R)-zP(z)}{2z(1-q)}(1-z)
\ee
and
\be
G^{ab,ab}({\bf r}={\bf e})=(1-q)\{ 2 \ \frac{3P(z_R)-2P(z)-1}{3z_R P(z_R)-zP(z)}-[1-q-2z\frac{P(z)}{3z_R P(z_R)-zP(z)}]\frac{P'(z)}
{zP'(z)+P(z)}\}
\ee
where we have used that $P({\bf r}={\bf e},z)=[P(z)-1]/z$. 

Given the appropriate boundary conditions, the set of Eqs. (52-55) allows to calculate $z$ and $z_R$ and thus the correlation functions 
and the Gibbs free energy for all values of $q$ and $\lambda$. However, from the definition of the Green's function, Eq. (45), one must have 
$0\leq z\leq 1$ and $0\leq z_R\leq 1$. In particular, 
$z=z_R=0$ corresponds to the high-field ($\Delta \rightarrow \infty$, $q \rightarrow 1$) or high-temperature ($\lambda \rightarrow 0$) 
limit, while $z=1$ and $z_R=1$ define respectively the spinodal line $\hat C^L_{22}({\bf k}={\bf 0})=0$ and
the limit of stability of the replica-symmetric solution in the $q-\lambda$ plane. At the freezing temperature $T_f$ when $\Delta=0$,
one has simultaneously $z(q=0)=z_R(q=0)=1$. Indeed, $c_{0,ref}^{ab,cd}(q=0)=0$, so that 
$z=z_R$ is always solution of Eq. (52) when $q=0$ (this is not the only solution, but the only one that is physically acceptable). 
Accordingly, $G_0^R(q=0)=G^L_{0,22}(q=0)$ 
so that $c_0^{ab,cd}(q=0)=0$, as required by symmetry (cf. the end of section 2).
 Another exact requirement is that $\Delta(q=0)=0$ for every $\lambda$ in the high-temperature phase.
Since $\Delta_{ref}(q=0)=0$ and $\partial \Delta/\partial \lambda=-q -(1/2) \partial G^{ab,ab}({\bf r}={\bf e})/\partial q$ (the integrated form
of Eq. (53)), this implies that 

\be
\left.\frac{\partial }{\partial q}G^{ab,ab}({\bf r}={\bf e})\right|_{q=0}=0 
\ee
in the high-temperature phase.We take this equation as the boundary condition for the PDE, Eq. (53), on the line $q=0$.

This suggests to take $G^{ab,ab}({\bf r}={\bf e})$ as the unknown function in the numerical integration of Eq. (53).
The PDE is then rewritten as a non-linear diffusion equation,

\be
\frac{\partial G}{\partial \lambda}
=-\frac{\left.\frac{\partial G}{\partial z}\right|_q }{\left.\frac{\partial C}{\partial z}\right|_q}\  (1 +\frac{1}{2}\frac{\partial ^2 G}{\partial q^2})
\ee
where $C \equiv \hat C^L_{22}({\bf k}={\bf 0})$ and $G \equiv G^{ab,ab}({\bf r}={\bf e})$. The integration is carried out by a simple explicit algorithm in which the ``space-like'' variable
$q$ and the ``time-like variable'' $\lambda$ are discretized, and the partial derivatives are approximated by finite difference 
representations \cite{PFTV1992}. The first derivative with respect to $\lambda$ is used to update $G$ 
at the temperature step $n+1$ by evaluating the second derivative with respect to $q$ at the step $n$. For given values of
$q$ and $G$, $z$ and $z_R$ are obtained from Eqs. (52) and (55) by using a standard Newton-Raphson 
algorithm (note that these equations does not depend explicitly on $\lambda$ and that $c_{0,ref}^{ab,cd}(q)$ can be tabulated once
and for all). $\hat C_{22}^L ({\bf k}={\bf 0})$ is then obtained from Eqs. (54).
Although an implicit method, such as the one used in Ref. \cite{PSD1998}, would have the advantage of being unconditionally stable, 
we have achieved numerical stability above and at $T_f$ by using small grid spacings, $\Delta q=5. 10^{-3}$ 
and $\Delta \lambda=10^{-5}$.

\section{Results and discussion}

As one decreases the temperature (i.e., increases $\lambda$)  in the numerical integration of the PDE, there is a value of $\lambda$
for which $z(q=0)=z_R(q=0)=1$. This corresponds to the divergence of the spin glass susceptibility, and it defines the 
freezing temperature $T_f$. The SCOZA results for $T_f$ are compared in Table 1 with
the predictions of a $1/d$ series expansion \cite{SF1988} for $d=9,7,5$ and with a recent Monte Carlo estimate \cite{MPRRZ1999}
for $d=3$ (we have only considered odd dimensions because the Green's function $P(z)$ is easier to evaluate). 
One notes a significant improvement over the predictions of mean-field theory ($T_f^{mf}=\sqrt 2d$). 
In fact, the results are
comparable to those of the Bethe-Peierls approximation \cite{SP1996}  for 
$d \geq 5$ and better for $d=3$. Note again that the present theory is exact in the replica-symmetric high-temperature phase 
when $d \rightarrow \infty$ and thus predicts the correct $T_f$ in this limit. It is likely
that the small deviations for $d=9$ and 7 come from the fact that only 
 $c_1^{ab,ab}(q,\beta)$ is nonzero in the simplified OZ approximation, Eq. (41). Indeed, the exact calculation of 
$C^{ab,ab}({\bf r}), C^{ab,ac}({\bf r})$ and $C^{ab,cd}({\bf r})$ to order $1/d$ shows that the three functions do
extend to nearest-neighbor separation \cite{RT1999}. But the introduction of  $c_1^{ab,ac}$ and $c_1^{ab,cd}$ 
would require self-consistency relations involving the three direct correlation functions separately. This is only possible if one 
breaks replica symmetry in some way. Below the upper critical dimension $d=6$, the fact that 
$\eta$ is no longer zero comes also into play 
and may explain the more significant deviations from the simulation results, especially for $d=3$ ($\eta \simeq\ -0.5$ 
according to Ref. \cite{BPC1996}).

When $J_0=0$, the zero-field internal energy per spin is given by $\beta u= - (\lambda/2) \  [1-G^{ab,ab}({\bf r}={\bf e})]$. Therefore,
at the freezing temperature, Eq. (55) readily gives $u_f/J=-c J/(2P(1)k_BT_f)$. The corresponding numerical 
values are also given in Table 1.
Finally, as an illustration of the behavior of the specific heat above $T_f$, we plot $c(T)=\partial u/\partial T$ for $d=5$ and $3$ in Fig. 1. 
Note the broad maximum 
that appears above the freezing temperature for $d=3$, in agreement with experimental and simulation data \cite{BY1986}. 
This is in contrast with the mean-field behavior that predicts a cusp at $T_f$.

In summary, the SCOZA results for the properties of the EA model at and above the freezing temperature as a function of dimensionality
are in reasonable agreement with the known estimates. This represents an encouraging step for proceeding further in studying
the stability of the replica-symmetric solution at temperatures below $T_f$. However, we have not yet been able to obtain a 
reliable solution of the SCOZA PDE below freezing and this work is still in progress.

\newpage

\appendix

\section{Direct correlation functions in the reference system}

From the expressions of the G's in the reference system and Eqs. (18) and (24) , one finds
\begin{mathletters}
\be
\hat {C}^L_{11,ref}=(1-4q+3r)D_{ref}
\ee
\be
\hat {C}^L_{22,ref}=(1-q)D_{ref}
\ee
\be
\hat {C}^L_{12,ref}=-\frac{1}{2}\hat {C}^L_{21,ref}=(m-t)D_{ref}
\ee
\end{mathletters}
where 
\bea
D_{ref}&=&\hat {C}^L_{11,ref}\hat {C}^L_{22,ref}+2(\hat {C}^L_{12,ref})^2\nonumber\\
&=&[(1-q)(1-4q+3r)+2(m-t)^2]^{-1} 
\eea
and $r$ and $t$ are given in Eqs. (29).
Moreover, Eq. (25) yields
 \bea
 \left(\begin{array}{ccr}
c^{ab}_{0,ref}\ \ &-\frac{1}{2}c^{a,bc}_{0,ref}\nonumber\\
2c^{a,bc}_{0,ref}-c^{a,ab}_{0,ref}\ \ &c^{ab,ac}_{ref}-\frac{3}{2}c^{ab,cd}_{0,ref}
\end{array} \right) =D_{ref}^2
 \left(\begin{array}{ccr}1-4q+3r\ \ &m-t\\
-2(m-t)\ \ &1-q\end{array} \right) \\ . \left(\begin{array}{ccr}
m^2-q\ \ &\frac{1}{2}(t-mq)\nonumber\\
m(1+q)-2t  \ \ &-q+\frac{1}{2}(3r-q^2)
\end{array} \right) . \ \left(\begin{array}{ccr}1-4q+3r\ \ &m-t\\
-2(m-t)\ \ &1-q\end{array} \right) \\
\eea
Finally, Eq. (22) for the replicon sector gives
\be
\hat {C}^R_{ref}=\frac{1}{1-2q+r}
\ee
It is now the straightforward to calculate all the $C_{ref}$'s from these equations. The results are

\bea
c^{aa}_{0,ref}&=&D_{ref}^2[(1-4q+3r)^2(1+m^2-2q)+2(m-t)(1-4q+3r)(2m+mq-3t)\nonumber\\
&+&(m-t)^2(6q+q^2-6r-1)]
\eea
\be
c^{ab}_{0,ref}=c^{aa}_{0,ref}-(1-4q+3r)D_{ref}
\ee
\bea
c^{a,bc}_{0,ref}&=&2\{(1-4q+3r)[(m(q-m^2)(1-q)-\frac{1}{2}(t-mq)(1+q-2m^2)]\nonumber\\
&+&(1-q)(m-t)(q+\frac{q^2}{2}-\frac{3r}{2})+(m-t)^2(2t-m-mq)\}D_{ref}^2
\eea
\be
c^{a,ab}_{0,ref}=c^{a,bc}_{0,ref}+(t-m)D_{ref}
\ee
\bea
c^{ab,cd}_{0,ref}&=&2 \{(1-q)^2(q+\frac{q^2}{2}-\frac{3r}{2})+(m-t)^2(1+2m^2-3q)+2(1-q)(m-t)(2t-m-mq)\}D_{ref}^2\nonumber\\
&+&2\frac{(1-q)(r-q)+(m-t)^2}{1-2q+r}D_{ref}
\eea
\be
c^{ab,ac}_{0,ref}=c^{ab,cd}_{0,ref}+\frac{(1-q)(r-q)+(m-t)^2}{1-2q+r}D_{ref}
\ee
\be
c^{ab,ab}_{0,ref}=2c^{ab,ac}_{0,ref}-c^{ab,cd}_{0,ref}+\frac{1}{1-2q+r}
\ee

\newpage

\begin{center}
\begin{tabular}{|c|c|c|c|}
\hline
\hspace{.25cm} {$d$} \hspace{2cm}& \hspace{.4cm} $k_BT_f^{scoza}/J$\hspace{.25cm}  & \hspace{.25cm} $k_BT_f^{be}/J$\hspace{.25cm}
& \hspace{.25cm} $u^{scoza}_f/J$\hspace{.25cm}\\
\hline
9 & 3.91 & $3.82^a$ & -2.16 \\
7 & 3.35 & $3.22^a$ & -1.91\\
5 & 2.66 & $2.41^a$  & -1.63\\
3 & 1.58 & $0.95 \pm 0.04^b$& -1.25\\
\hline
\end{tabular}

\quad
\quad
\end{center}
{Table 1: Freezing temperature and internal energy per spin at $T_f$ of the d-dimensional EA spin glass model 
with Gaussian couplings and $J_0=0$.
The SCOZA results for $T_f$ are compared 
with the best series or simulation estimates. a: Ref. \cite{SF1988}. b: Ref. \cite{MPRRZ1999}}.
For $d=9,7,5$, $T_f$  is obtained from the $1/d$ series estimates of the freezing temperature for the $\pm J$ distribution 
using the transformation formula given in Ref. \cite{SF1988}.

\large

\vspace*{1.cm}
\begin{figure}[h]
\vspace*{0.5cm}
\hspace*{2.0cm}
\leavevmode
\epsfxsize= 80pt
\epsffile[ 100 370 200 670]{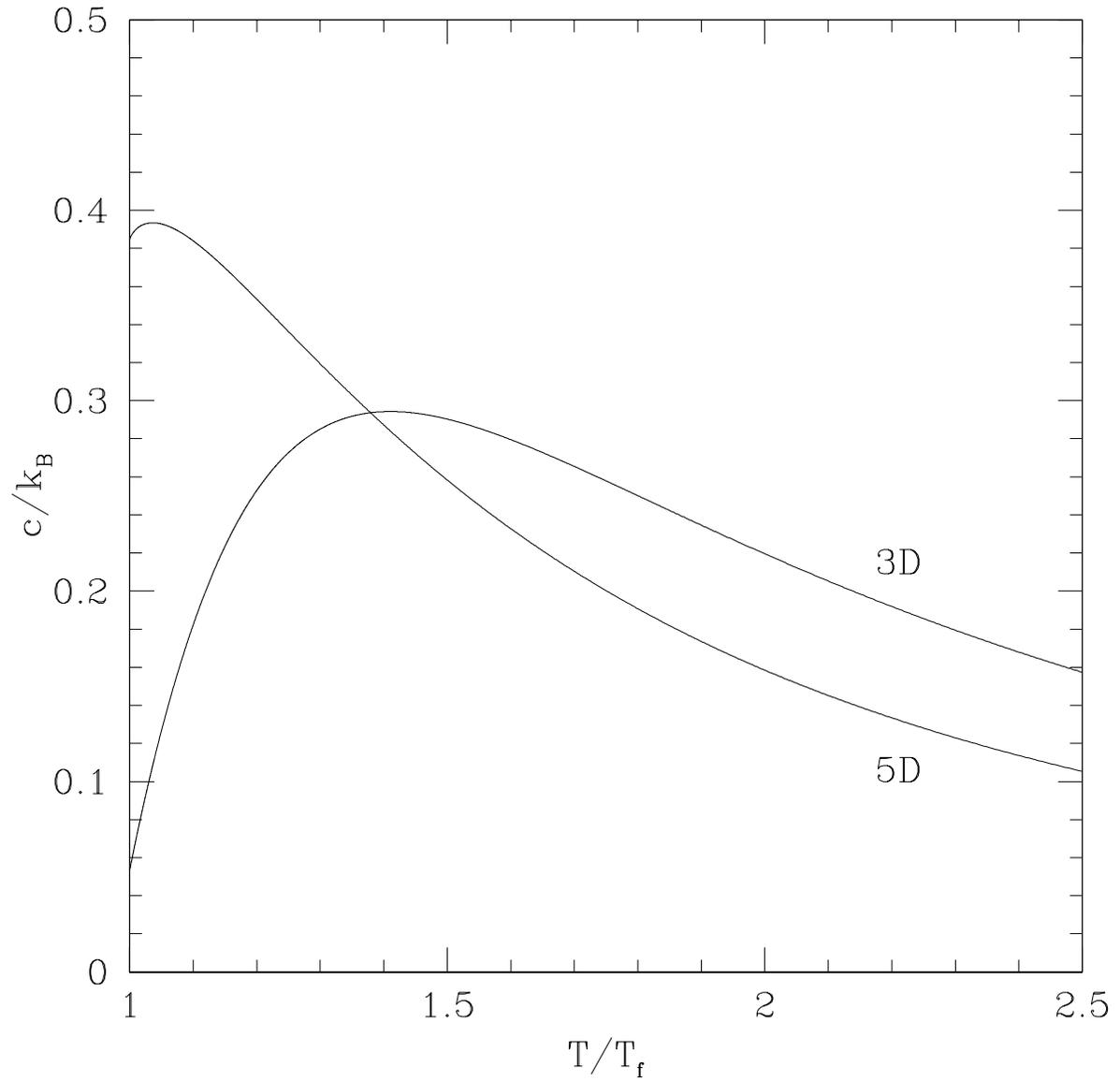}

\vspace*{7.0cm}
\caption{SCOZA predictions for the specific heat vs reduced temperature $T/T_f$ of the EA spin glass model 
with Gaussian couplings and $J_0=0$ for  $T\geq T_f$.}

\end{figure}

\end{document}